\documentclass[aps,nofootinbib,superscriptaddress,10pt,floatfix]{revtex4}
\usepackage{mathrsfs}
\usepackage{graphicx}
\usepackage{amsmath,amssymb}
\usepackage{hyperref}
\usepackage{braket}
\usepackage{subfigure}
\usepackage{subfloat}
\usepackage{float}
\usepackage[usenames]{color}
\usepackage[normalem]{ulem}
\usepackage{comment}

\hypersetup{
    colorlinks=true,
    linkcolor=black,
    citecolor=blue,
}

\allowdisplaybreaks

\def\be{\begin{equation}}
\def\ee{\end{equation}}
\def\ba{\begin{eqnarray}}
\def\ea{\end{eqnarray}}

\newcommand{\bn}{\mathbf{n}}
\newcommand{\HH}{\mathcal{H}}
\newcommand{\bV}{\mathbf{V}}
\newcommand{\bk}{\mathbf{k}}
\newcommand{\bx}{\mathbf{x}}
\newcommand{\theg}{\theta_{g}}
\newcommand{\B}{{\rm B}}
\newcommand{\F}{{\rm F}}

\begin{document}

\title{Modified Einstein versus Modified Euler for Dark Matter}
\author{Camille Bonvin}
\email[]{camille.bonvin@unige.ch}
\affiliation{Departement de Physique Theorique and Center for Astroparticle Physics, Universite de Geneve, Quai E. Ansermet 24, CH-1211 Geneve 4, Switzerland}
	
\author{Levon Pogosian}
\email[]{levon@sfu.ca}
\affiliation{Department of Physics, Simon Fraser University, Burnaby, BC, V5A 1S6, Canada}

\begin{abstract}
Modifications of General Relativity generically contain additional degrees of freedom that can mediate forces between matter particles. One of the common manifestations of a fifth force in alternative gravity theories is a difference between the gravitational potentials felt by relativistic and non-relativistic particles, also known as ``the gravitational slip''. In contrast, a fifth force between dark matter particles, due to dark sector interaction, does not cause a gravitational slip, making the latter a possible smoking gun of modified gravity. In this article, we point out that a force acting on dark matter particles, as in models of coupled quintessence, would also manifest itself as a measurement of an \emph{effective} gravitational slip by cosmological surveys of large-scale structure. This is linked to the fact that redshift-space distortions due to peculiar motion of galaxies do not provide a measurement of the true gravitational potential if dark matter is affected by a fifth force. Hence, it is extremely challenging to distinguish a dark sector interaction from a modification of gravity with cosmological data alone. Future observations of gravitational redshift from galaxy surveys can help to break the degeneracy between these possibilities, by providing a direct measurement of the distortion of time. We discuss this and other possible ways to resolve this important question.
\end{abstract}
	
\maketitle

\section{Modified gravity vs dark force}	

The discovery of cosmic acceleration~\cite{Perlmutter:1998np,Riess:1998cb} and the unknown nature of dark matter (DM) prompted extensive studies of modified gravity theories. Generically~\cite{Lovelock:1971yv,Lovelock:1972vz}, such theories involve, in addition to the metric tensor, new dynamical degrees of freedom, with a scalar field being the most commonly studied example~\cite{Horndeski:1974wa,Deffayet:2011gz}. In these scalar-tensor theories, gravitational attraction between matter particles is mediated by the curvature of spacetime as well as the scalar field.
At the level of linear cosmological perturbations, this ``fifth force'' not only enhances the rate of gravitational clustering of matter, but manifests itself as a non-zero ``gravitational slip'' \cite{Amendola:2007rr,Daniel:2008et}, namely, a difference between the Newtonian potential $\Psi$ and the curvature perturbation $\Phi$. One can search for evidence of $\Psi \ne \Phi$ by combining observations of galaxy redshift-space distortions (RSD) and weak gravitational lensing (WL), along with other cosmological data \cite{Zhang:2007nk,Song:2010fg,Amendola:2012ky}. A measurement of $\Phi \ne \Psi$ is often considered to be the smoking gun of modified gravity.

What if instead of having modifications of gravity affecting all matter, only the DM particles experience an attractive force due to some non-gravitational dark sector interaction? Can cosmological observations distinguish a dark sector force, that affects only DM, from a modification of gravity that alters gravity for all matter? Phrasing it in mathematical terms, can one distinguish a modification of the Einstein equations from a modification of the Euler equation for DM? While finding any evidence of a fifth force would be of profound importance by itself, knowing whether it is of gravitational or particle origin is an equally fundamental question.

This question is not new and has been discussed, for example, in the context of scalar field dark energy \cite{EuclidTheoryWorkingGroup:2012gxx,Amendola:2016saw}. A minimally coupled scalar field is usually referred to as quintessence \cite{Wetterich:1994bg,Zlatev:1998tr}, while a scalar field coupled only to DM would be classified as coupled quintessence (CQ). (Note that in the earlier literature, {\it e.g.} \cite{Amendola:1999er}, the term CQ was also used to refer to coupling to all matter, but in more recent years CQ was generally used to refer to the DM-only coupled case \cite{Amendola:2016saw,Barros:2018efl}.) 
On the other hand, a scalar field universally coupled to all matter would be referred to as a scalar-tensor theory~\cite{EuclidTheoryWorkingGroup:2012gxx,Amendola:2016saw} and, hence, considered to be modified gravity. Several publications suggested that a way to differentiate between CQ and scalar-tensor gravity would be to measure the gravitational slip \cite{EuclidTheoryWorkingGroup:2012gxx,Song:2010rm,Motta:2013cwa,Amendola:2016saw}. 
This expectation, however, relies on our ability to measure the perturbation of the velocity field of the normal matter (``baryons''), and use it to infer the underlying large-scale  $\Psi$.

In this paper we argue that this is not possible with current observations. The reason is that the baryons  we observe are confined in galaxies and clusters. As such their velocity is linked to the velocity of galaxies and, therefore, they do not trace the large-scale $\Psi$, if DM experiences a fifth force. The effective Newtonian potential inferred from RSD, when compared with WL measurements, would consequently yield a non-zero measured gravitational slip indistinguishable from that coming from modified gravity.

Fortunately, the next generation of large-scale structure surveys has the potential to break this degeneracy between modified gravity and a dark force acting on DM (hereafter called dark force), by providing a measurement of the distortion of time. This novel observable has the advantage of being directly sensitive to $\Psi$, even in the presence of a dark force.

\section{The smoking gun argument}
\label{sec:argument}

We start by comparing two models: a scalar-tensor theory of generalized Brans-Dicke (GBD) type, and a CQ model. While the equations of motion and the perturbations we show are specific to these two models, the argument is general and holds for any modified gravity theory and dark force model.

The action for GBD takes the form
\be
S^{\rm GBD} = \int \mathrm{d}^4 \sqrt{-g} \left[ \frac{A^{-2}(\phi)}{16\pi G}R - \frac{1}{2} \partial_{\mu} \phi \, \partial^{\mu} \phi - V(\phi) + \mathcal{L}_{\mathrm{m}} (\psi_\mathrm{DM}, \psi_\mathrm{SM}, g_{\mu \nu}) \right],
\label{eq:action_st}
\ee
where $G$ is the Newton's constant, $R$ is the Ricci scalar built from $g_{\mu \nu}$ and its derivatives, $g$ is the metric determinant, $A$ is a generic function of the scalar field $\phi$ and $V$ is its potential. $\mathcal{L}_{\mathrm{m}} (\psi_\mathrm{DM}, \psi_\mathrm{SM}, g_{\mu \nu})$ is the Lagrangian density of all matter that includes the Standard Model (SM) particle fields, collectively denoted as $\psi_\mathrm{SM}$, and the DM particles, denoted as $\psi_\mathrm{DM}$, with both following the geodesics of the metric $g_{\mu \nu}$. \emph{Throughout this paper, $g_{\mu \nu}$ denotes the metric of the ``baryon frame''}, {\it i.e.}\ the metric whose geodesics are followed by the SM particles (which, in the case of the scalar-tensor theories, is the same for baryons and DM).

Let us compare the GBD action (\ref{eq:action_st}) to the action of CQ, with the scalar field conformally coupled only to DM,
\be
S^{\rm CQ} = \int \mathrm{d}^4 \sqrt{-g} \left[ \frac{1}{16\pi G}R - \frac{1}{2} \partial_{\mu} \phi \, \partial^{\mu} \phi - V(\phi) + \mathcal{L}_{\mathrm{SM}} (\psi_\mathrm{SM}, g_{\mu \nu}) + \mathcal{L}_{\mathrm{DM}} (\psi_\mathrm{DM}, A^{2} (\phi) g_{\mu \nu})  \right],
\label{eq:action_cq}
\ee
in which the gravitational part of the action is not modified in the baryon frame $g_{\mu \nu}$, and with DM following geodesics of $A^{2} (\phi) g_{\mu \nu}$. 

We always interpret the observations in the ``baryon frame'', in which the masses of the SM particles are constant. With that in mind, let us compare the equations governing linear cosmological perturbations in GBD and CQ theories. We work with the linearly perturbed flat Friedmann-Lema\^itre-Robertson-Walker (FLRW) metric in the conformal Newtonian gauge, with the line element given by
\be
ds^2 = g_{\mu \nu} dx^\mu dx^\nu = a^2(\tau) \left[ -(1 + 2\Psi) d\tau^2 + (1-2\Phi) d \bx^2 \right]\, ,
\label{eq:metric}
\ee
where $\tau$ denotes conformal time and $a$ is the scale factor.
Neglecting radiation, the relevant variables are $\Psi$, $\Phi$, the baryon and (Cold) DM density contrasts, $\delta_{b} = \delta \rho_{b}/\rho_{b}$ and $\delta_{c} = \delta \rho_{c}/\rho_{c}$, and their velocity divergences, $\theta_{b}$ and $\theta_{c}$.
As shown in Section~\ref{sec:GBDCQ}, in both GBD and CQ, the equations governing the evolution of these variables can be combined into an evolution equation for the matter density contrast,
\begin{align}
\ddot{\delta}+ {\cal H} \dot{\delta} = 4 \pi G_{\rm eff} a^2 \rho\,\delta\, , \label{eq:delta_evol}
\end{align}
where the over-dots denote derivatives with respect to 
$\tau$, ${\cal H}$ is the Hubble parameter in conformal time, $\rho\delta=\rho_c \delta_c+\rho_b\delta_b$, and
$G_{\rm eff}$ is the effective gravitational coupling that takes the following forms,
\begin{align}
G^{\rm GBD}_{\rm eff} = G\left[ 1+{2 \tilde{\beta}^2k^2\over a^2m^2 + k^2}\right] \quad\quad\mbox{and}\quad\quad G^{\rm CQ}_{\rm eff}  =G\left[ 1 + {2 \tilde{\beta}^2k^2\over a^2m^2 + k^2} \left( \rho_c \over \rho \right)^2 \left( \delta_c \over \delta \right)\right]\, , \label{eq:Geff}
\end{align}
where $\tilde{\beta}^2 = \beta^2/ 8\pi G$, $\beta=A,_{\phi}/A$ is the scalar field coupling strength and $m^2$ is the effective mass that sets the range of the fifth force.
We see that the effective gravitational couplings are very similar in the two models. The only difference is a small suppression of the impact of the fifth force in $G^{\rm CQ}_{\rm eff}$, due to the fact that $\sim 15\%$ of matter does not feel the fifth force. This difference is, however, degenerate with the unknown coupling $\tilde{\beta}$. We see, therefore, that GBD and CQ are impossible to distinguish through the growth of structure alone. An observer looking for departures from $\Lambda$CDM by fitting $G_{\rm eff}$ to the galaxy growth data ({\it e.g.}\ using \textsc{MGCAMB} \cite{Zucca:2019xhg}) 
would measure a $G_{\rm eff}>1$ either way. 
Note that the argument derived here in the case of a scalar field holds in general: modifications to Poisson equation (due to modified gravity) and modifications to Euler equation (due to a dark fifth force) are generically indistinguishable at the level of the growth 
rate~\cite{Castello:2022uuu}, which is the quantity measured by RSD.

On the other hand, the two types of theories differ at the level of the gravitational potentials. In GBD 
the two potentials differ, $\Phi \ne \Psi$, hence $\eta \equiv \Phi/\Psi \ne 1$, whereas in CQ the Einstein equations are not modified, and therefore, at late times, $\eta=1$. This suggests that one could differentiate the two cases by measuring $\eta$~\cite{Motta:2013cwa,Amendola:2016saw}, making it a smoking gun for modified gravity. Note that modified gravity effects on linear perturbations can, in principle, be mimicked by a dark fluid with appropriately tuned state functions (see {\it e.g.} \cite{Kunz:2006ca,Battye:2012eu}). Here, rather than aiming to distinguish between a modified gravity and a hypothetical fluid, we compare a modified gravity in which a fifth force affects all matter with a theory in which the same type of force acts only on DM, with no additional dark ingredients.

In practice, deviations from GR are often parameterized with two functions $\mu$ and $\Sigma$ that depend on $a$ and the wavenumber $k$
\ba
k^2 \Psi &=& -4 \pi \mu(a,k) G a^2 \rho\,\delta\, , \\
k^2 (\Phi + \Psi) &=& -8 \pi \Sigma(a,k) G a^2 \rho\,\delta \ , \label{eq:Sigma_def}
\ea
where, in GBD, 
\be
\label{eq:mu_st}
\mu=\frac{G_{\rm eff}^{\rm GBD}}{G} \quad\mbox{and}\quad \Sigma = {1\over 2}\mu (1+\eta) = 1 \, ,
\ee
while in CQ, $\mu=\Sigma=\eta=1$. In theory, combining a measurement of baryon velocities, determined by the Newtonian potential $\Psi$, with 
WL data that measure $\Phi + \Psi$, would yield a measurement of both $\mu$ and $\Sigma$ and, therefore, determine $\eta$. However, as we show below, this test would not work in practice because the baryons we observe are confined to galaxies and, hence, move together with the galactic DM. This means an observer would measure an effective $\eta^{\rm fit} \ne 1$ even if there is no intrinsic gravitational slip.

\section{The observed gravitational slip}

\begin{figure}
  \includegraphics[width=0.6\linewidth]{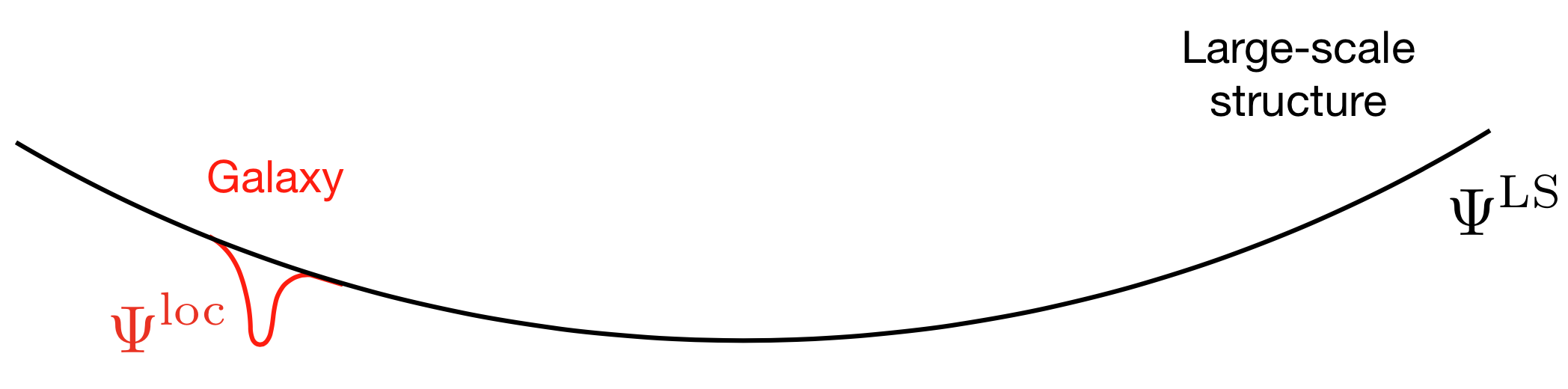}
  \caption{Illustration of the large-scale and local contribution of the gravitational potential $\Psi$. 
  }
  \label{fig:Psi}
\end{figure}

To understand how the gravitational slip is measured from RSD and 
WL, let us first review how these observables are constructed. Redshift surveys map the distribution of galaxies and measure the fluctuation in the galaxy number counts, given in Fourier space by
\begin{align}
\Delta(\bk, z)&=\delta_g(\bk, z)-\frac{1}{\HH}\mu_k^2\,\theta_b(\bk, z)\, , \label{eq:Deltak_main}
\end{align}
where $z$ is the redshift, $\mu_k=\hat{\bk}\cdot\bn$,
and $\bn$ is the direction of observation (considered fixed in the flat-sky approximation). The first term is the intrinsic fluctuation in the distribution of galaxies $\delta_g$, related to the (total) matter density contrast through the bias $\delta_g=b\,\delta$. The second term 
is due to RSD~\cite{Kaiser:1987qv}, accounting for the fact that the redshift of the galaxies is affected by the peculiar velocity of the baryons (from which the light that we receive is emitted) with respect to us. As shown in Section~\ref{sec:vbaryon}, the velocity of baryons can be decomposed into two terms: the velocity of the baryons with respect to the center of mass of the galaxy, and the galactic center of mass velocity with respect to the Hubble flow. 
These two terms are sensitive to different ingredients. As illustrated in Fig.~1, the velocity of the baryons with respect to the center of mass is governed by the local gravitational potential of the galaxy, whereas the velocity of the center of mass is driven by the large-scale gravitational potential. Since RSD surveys measure correlations of galaxy number counts at large separations (well above the size of a galaxy), the first velocity contribution vanishes, since it is not correlated on large scales. Consequently, the RSD power spectrum is only affected by the motion of the galactic center of mass, and we can effectively replace $\theta_b$ in Eq.~\eqref{eq:Deltak_main} by the center of mass velocity, denoted by $\theta_g$. In GBD, the center of mass moves according to the large-scale gravitational potential $\Psi^{\rm LS}$. In the CQ model however, the center of mass velocity is also affected by the fifth force:
\begin{align}
&\mbox{GBD}:\quad\quad \dot{\theta}_g+\HH\theg=k^2\Psi^{\rm LS}\, ,\label{eq:VgGBD}\\
&\mbox{CQ}:\quad\quad\dot{\theta}_g+\HH\theg=k^2\Psi^{\rm LS}+\frac{\rho_c}{\rho}k^2\beta\delta\phi\equiv k^2\Psi^{\rm eff}\, .\label{eq:VgCQ}
\end{align}
Therefore, we see that in the CQ case, RSD do not allow us to reconstruct the large-scale gravitational potential $\Psi^{\rm LS}$, even though the fifth force does not act directly on baryons.

To link this to standard RSD analyses, we relate the galaxy velocity to the matter density contrast, assuming that the continuity equation is valid in both models (see Section~\ref{sec:vbaryon}). With this, the RSD power spectrum becomes
\begin{align}
P^{\rm gal}(k,\mu_k,z)&=\left(b^2+\mu_k^2f\right)^2P_{\delta\delta}(k,z)\, , \label{eq:Pf}
\end{align}
where $f\equiv d\ln\delta / d\ln a$ is the growth rate and $P_{\delta\delta}$ is the matter power spectrum.
Both $f$ and $P_{\delta\delta}$ are determined by the solution to Eq.~\eqref{eq:delta_evol} and, therefore, directly affected by $G_{\rm eff}$ that has similar forms in GBD and CQ (see Eq.~\eqref{eq:Geff}).

The second relevant observable is WL, measured through cosmic shear or lensing of the cosmic microwave background. The WL convergence, $\kappa$, 
probes the sum of the two gravitational potentials via
\begin{align}
\kappa(\bn, z) & =  \int_0^{r(z)}d r'\frac{r(z)-r'}{2 r(z)r'}\,\Delta_\Omega(\Phi+\Psi)\big(\bn, r'\big)\, ,  \label{eq:kappa}
\end{align}
where $r$ is the comoving distance to the sources and $\Delta_\Omega$ is the Laplace operator on the sphere. As for RSD, the correlations of 
convergence over large distances are only affected by the large-scale part of the potentials. Lensing correlations, therefore, effectively provide a measurement of the power spectrum of $\Phi^{\rm LS}+\Psi^{\rm LS}$, which can be related to 
$P_{\delta\delta}$ through Eq.~\eqref{eq:Sigma_def}, 
\begin{align}
    P^{(\Phi+\Psi)}(k,z)
    =9H_0^4\Omega_m^2(1+z)^2\Sigma^2(k,z)P_{\delta\delta}(k,z)\, , \label{eq:Plensing}
\end{align}
where $H_0$ is the Hubble parameter today and $\Omega_m$ is the matter density parameter.
WL measurements are therefore sensitive to two ingredients: the parameter $\Sigma$, which links the gravitational potentials to the density fluctuation, and the effective gravitational coupling $G_{\rm eff}$, which affects the density power spectrum $P_{\delta\delta}$. 

From Eqs.~\eqref{eq:Pf} and~\eqref{eq:Plensing}, we see that combining WL with RSD allows one to measure both $G_{\rm eff}$ and $\Sigma$ simultaneously. From those, we can infer $\mu^{\rm fit}$ and $\eta^{\rm fit}$ that one would obtain under the assumption that Euler equation is unmodified. For GBD, we have
\begin{align}
    \mu^{\rm fit}&=\frac{G_{\rm eff}^{\rm GBD}}{G}=\mu^{\rm GBD}>1\, ,\\
    \eta^{\rm fit}&=\frac{2\Sigma^{\rm fit}}{\mu^{\rm fit}}-1=\frac{2}{\mu^{\rm fit}}-1=\eta^{\rm GBD}<1 \, ,
\end{align}
{\it i.e.}, we would observe a non-zero gravitational slip, $\eta^{\rm fit}<1$, as expected. 
For CQ, we have
\begin{align}
    \mu^{\rm fit}&=\frac{G_{\rm eff}^{\rm CQ}}{G}>1\,\\
    \eta^{\rm fit}&=\frac{2\Sigma^{\rm fit}}{\mu^{\rm fit}}-1=\frac{2}{\mu^{\rm fit}}-1<1\, .
\end{align}
Hence, even though the gravitational slip is zero in CQ, one would still measure $\eta^{\rm fit}<1$ by combining RSD with WL.
This clearly demonstrates that measuring $\eta \ne 1$ from RSD and WL \emph{is not a smoking gun} for modified gravity -- it can also be due to a fifth force acting solely on dark matter. 

\begin{figure}
  \includegraphics[width=0.5\linewidth]{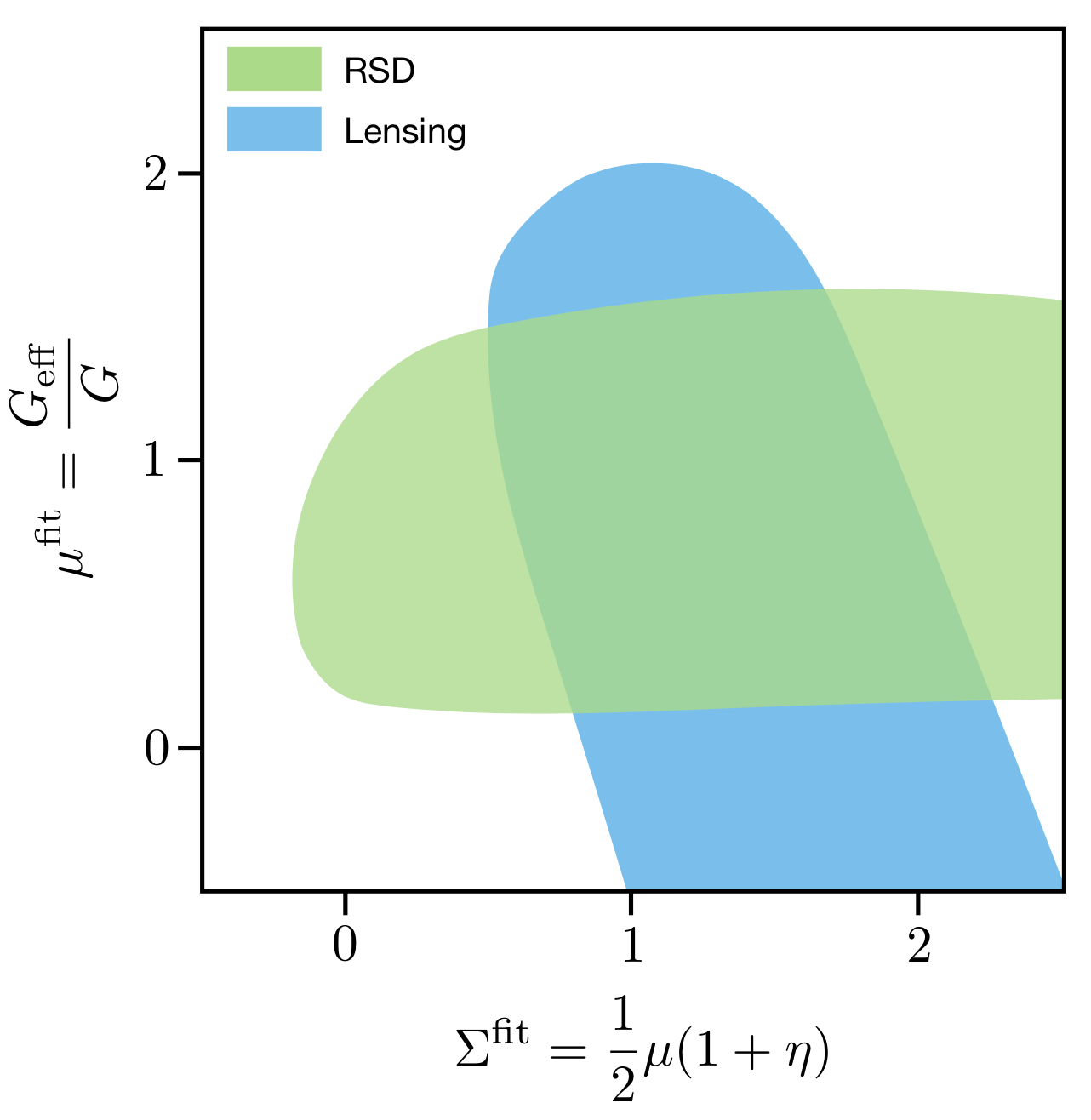}
  \caption{Illustration of the constraints on $\mu$ and $\Sigma$ from combined measurements of RSD and WL. The observed $\mu^{\rm fit}$ is related to $G_{\rm eff}$, hence affected by a dark fifth force. The observed $\Sigma^{\rm fit}$ on the other hand is related to the true $\mu$, i.e.\ the one that enters into Poisson equation, and which is exactly equal to one in models with a dark force. 
  }
  \label{fig:musigma}
\end{figure}

While we used CQ as our example, the effective gravitational slip is present in any model that breaks the weak equivalence principle for dark matter, {\it i.e.} any model where a dark force is acting solely on DM. As is schematically shown in Fig.~2, RSD provide constraints on $\mu^{\rm fit}$ (green region), whereas WL constrains both $\mu^{\rm fit}$ and $\Sigma^{\rm fit}$ (blue region). Since lensing probes the geometry of the Universe, $\Sigma^{\rm fit}$ is always equal to the true $\Sigma$ entering in Eq.~\eqref{eq:Plensing}. Therefore, even if there is a dark fifth force, $\Sigma^{\rm fit}$ is related to the \emph{true} $\eta$ and $\mu$. On the other hand, $\mu^{\rm fit}$ is fitted using the evolution equation for the density contrast, which depends on $G_{\rm eff}$. Therefore, if there is a dark fifth force, $\mu^{\rm fit}$ differs from the true $\mu$. As a consequence, when combining $\Sigma^{\rm fit}=\mu(1+\eta)/2=1$ with $\mu^{\rm fit}> 1$ in models with a dark force, we automatically obtain $\eta^{\rm fit}< 1$.

In~\cite{Motta:2013cwa}, it was argued that this problem could be circumvented by using RSD to measure directly the Newtonian potential $\Psi$, instead of constraining $G_{\rm eff}$ (and hence $\mu^{\rm fit}$) through the growth rate. However, since the RSD power spectrum is governed by the galaxy center of mass, $\theta_g$, which is affected by the effective gravitational potential $\Psi^{\rm eff}$ (see Eq.~\eqref{eq:VgCQ}), this method would also lead to a measurement of $\eta^{\rm fit}< 1$ (see Section~\ref{sec:vbaryon} for a detailed derivation).

\section{Distinguishing modified gravity from a dark force with gravitational redshift}
\label{sec:dipole}

Fortunately, the coming generation of galaxy surveys will allow us to measure a new observable, \emph{gravitational redshift}, which can be used to unambiguously distinguish between a dark fifth force and a modification of gravity.

As explained above, the main problem with measuring $\eta$ from RSD and WL is that RSD are not a tracer of the true large-scale gravitational potential, $\Psi^{\rm LS}$, if DM is affected by a fifth force. However, there are other distortions contributing to the observed galaxy number counts $\Delta$~\cite{Bonvin2011,Yoo:2009au,Challinor:2011bk}. Among these effects, one is particularly important for testing gravity: the effect of gravitational redshift. This effect encodes the fact that, when light escapes a gravitational potential, its energy is shifted to the red. Contrary to WL, which is sensitive to the sum of the two gravitational potentials (both time and space distortions deviate the trajectory of light), the shift in energy is only due to the time distortion. Therefore, gravitational redshift provides a measurement of the true $\Psi$, even in the presence of a fifth force.
Combining this with WL will allow us to measure the true gravitational slip and, consequently, distinguish a dark fifth force from a modification of gravity.

In practice, the gravitational redshift contribution to 
$\Delta$ is very small, and contributes in a negligible way to standard analyses.
However, this effect has the specificity to generate asymmetries in the distribution of galaxies~\cite{Bonvin:2013ogt}. For this reason, it was proposed to isolate it by searching for asymmetries in the cross-correlation of two populations of galaxies, for example a bright (B) and faint (F) population~\cite{McDonald:2009ud,Croft:2013taa,Bonvin:2013ogt}. Gravitational redshift is, however, not the only contribution that generates asymmetries in the correlation function: there are also Doppler effects, proportional to the galaxy center of mass velocity, that have the same property~\cite{Yoo:2012se,Bonvin:2013ogt}. Any measured asymmetry will, therefore, be due to a combination of these Doppler terms and gravitational redshift. These terms are generally called relativistic effects in the literature, even though, in reality, only gravitational redshift is a pure effect of general relativity. They contribute to the galaxy number counts as:
\begin{align}
\Delta^{\rm rel}&(\bk, z)=i \mu_k\left[-\frac{k}{\HH}\Psi(\bk, z)+\left(1-5s+\frac{5s-2}{\mathcal H r}-\frac{{\dot{\HH}}}{\mathcal H^2}+f^{\rm evol}\right) \frac{\theta_g(\bk,z)}{k}+\frac{\dot{\theta}_g(\bk, z)}{k\HH}\right]\, ,\nonumber
\end{align}
where $s$ is the magnification bias and $f^{\rm evol}$ is the evolution bias. Contrary to RSD, these relativistic effects generate contributions to the galaxy power spectrum with odd powers of $\mu_k$, and can be isolated by looking for a dipole and octupole. The dipole, which is the dominant contribution, is given by
\begin{align}
P_{\rm BF}^{(1)}(k,z)&= i\,\alpha\left(f,\dot{f},\Theta_\B, \Theta_\F \right)\frac{\HH}{k}P_{\delta\delta}(k,z) + i(b_\B-b_\F)\frac{k}{\HH}P_{\delta\Psi^{\rm LS}}(k,z)\, ,\label{eq:dip}    
\end{align}
where $\alpha$ is a function of the growth rate $f$ and its time derivative, as well as of $\Theta_\B$ and $\Theta_\F$ that encode the dependence of the dipole on the bias, magnification bias and evolution bias of the bright and faint population, respectively. The dipole is suppressed by one power of $\mathcal{H}/k$ with respect to the even multipoles (see 
Section~\ref{sec:multipoles}), and it is consequently too small to be measured in current surveys~\cite{Gaztanaga:2015jrs}. However,  forecasts have shown that it will be detectable with high significance with the coming generation of surveys, like DESI and the Square Kilometer Array (SKA2)~\cite{Beutler:2020evf,Bonvin:2018ckp}. 

From Eq.~\eqref{eq:dip}, we see that combining the dipole with the even multipoles (that depend on $P_{\delta \delta}$) allows one to directly measure $P_{\delta\Psi^{\rm LS}}(k,z)$~\cite{Sobral-Blanco:2021cks,Sobral-Blanco:2022oel},
that can be used to unambiguously distinguish between modified gravity and a dark fifth force. In practice, this can be done in two complementary ways. The first possibility is to look directly for modifications of gravity by combining $P_{\delta\Psi^{\rm LS}}(k,z)$ with galaxy-galaxy lensing (see {\it e.g.}~\cite{DES:2021qnp}), which measures the correlation of density with lensing: $P_{\delta(\Phi^{\rm LS}+\Psi^{\rm LS})}(k,z)$. The ratio of these two measured quantities 
gives $\eta$:
\begin{align}
    \frac{P_{\delta(\Phi^{\rm LS}+\Psi^{\rm LS})}(k,z)}{P_{\delta\Psi^{\rm LS}}(k,z)}=1+\eta(k,z)\, .~\label{eq:etatrue}
\end{align}
In~\cite{Tutusaus:2022cab} it was shown that, with this method, $\eta$ can be measured with a precision of 20-30\% at low redshift (in 4 bins, between $z=0.2$ to $z=0.7$), by combining spectroscopic measurements from SKA2 and photometric measurements from Vera Rubin Observatory~\cite{LSST:2008ijt}.
Since the denominator of Eq.~\eqref{eq:etatrue} depends on the true gravitational potential, a detection of $\eta\neq 1$ with this method would truly be a smoking gun for modified gravity. Models with a dark fifth force would give $\eta=1$.

The second way of using $P_{\delta\Psi^{\rm LS}}(k,z)$ to distinguish between modified gravity and a dark fifth force is to combine it with RSD to directly test the validity of the weak equivalence principle, {\it i.e.}\ to constrain the strength of the fifth force~\cite{Bonvin:2018ckp}. More precisely, one can compare $P_{\delta\Psi^{\rm LS}}(k,z)$ with $P_{\delta \theta_g}(k,z)$ measured from RSD, to directly probe Euler equation for galaxies in~Eqs.~\eqref{eq:VgGBD} and~\eqref{eq:VgCQ}, and measure the fifth force, proportional to $\beta$ in the case of CQ. In~\cite{Castello:2022uuu}, it was shown that, with this method, modifications of Euler equation can be constrained and disentangled from a change in the Poisson equation at the level of 15\%, with SKA2. Note that these forecasts were based on a particular parameterization in which modifications were proportional to the dark energy density fraction, as commonly assumed in other literature~\cite{Ade:2015rim}.
The constraints would be tighter in models where deviations could occur at earlier epochs.

\section{Conclusions}

Current data are not able to distinguish unambiguously between modifications to Einstein equations and modifications to Euler equation. The limitation is due to the fact that large scale structure is described by four fields: $\delta_g$, $\theta_g, \Phi$ and $\Psi$, whereas current observations can only measure three quantities: $\delta_g$, $\theta_g$ and $\Phi+\Psi$. 
Measuring the galaxy dipole with future surveys will add the missing information, allowing one to differentiate between a dark fifth force and a modification of gravity.

\section{Methods}
\label{sec:methods}

\subsection{Effective gravitational couplings in GBD and CQ}
\label{sec:GBDCQ}

To derive Eqs.~(\ref{eq:delta_evol}) and (\ref{eq:Geff}), for simplicity, we will adopt the quasi-static approximation (QSA), in which one restricts to sub-horizon scales and assumes that the time derivatives of the metric and the scalar field perturbations are much smaller than their spatial derivatives. Under the QSA, in Fourier space, the relevant equations in the baryon frame are

%\vspace{0.2cm}

\hspace{-0.4cm}\fbox{
\begin{minipage}{0.424\linewidth}\vspace{0pt}%
 %       \fontsize{10}{6}\selectfont 
 Generalized Brans-Dicke (GBD)
%\ba
\begin{align}
&k^2 \Phi = -4 \pi Ga^2 \left(\rho_{b} \delta_{b} + \rho_{c} \delta_{c}\right) - \beta k^2 \delta \phi \label{eq:poisson_st} \\
&k^2 (\Phi - \Psi) = -2 \beta k^2 \delta \phi \label{eq:slip} \\
&\dot{\delta}_{b}+\theta_{b} = 0 \label{eq:STcont_b} \\
&\dot{\theta}_{b}+{\cal H} \theta_{b} = k^{2} \Psi \label{eq:STeuler_b}\\
&\dot{\delta}_{c}+\theta_{c} = 0 \label{eq:STcont_c}\\
&\dot{\theta}_{c}+{\cal H} \theta_{c} = k^{2} \Psi \label{eq:STeuler_c}\\
\label{eq:dphi}
&\delta \phi = - \frac{\beta (\rho_{c} \delta_{c} + \rho_{b} \delta_{b})}{m^2 + k^2/a^2} \\
&\square \phi = V_{, \phi}+\beta({\rho_{c}+\rho_b}) \equiv V^{\rm eff},_{\phi} \label{eq:Veff_st}\\
\label{eq:delta_ddot_st}
&\ddot{\delta}+ {\cal H} \dot{\delta} = 4 \pi G a^2 \rho\delta\left[ 1 + {2 \tilde{\beta}^2k^2\over a^2m^2 + k^2} \right]
\end{align}
%\ea
\end{minipage}%
    }%
\fbox{%
\begin{minipage}{0.549\linewidth}\vspace{0pt}%
%        \fontsize{10}{6}\selectfont 
Coupled Quintessence (CQ)
%\ba
\begin{align}
&k^2 \Phi = -4 \pi Ga^2 \left(\rho_{b} \delta_{b} + \rho_{c} \delta_{c} \right) \\
&k^2 (\Phi - \Psi) = 0 \\
&\dot{\delta}_{b}+\theta_{b} = 0 \label{eq:CQcont_b} \\
&\dot{\theta}_{b}+ {\cal H} \theta_{b} = k^{2} \Psi \label{eq:CQeuler_b} \\
&\dot{\delta}_{c}+\theta_{c} = 0 \label{eq:CQcont_c}\\
&\dot{\theta}_{c}+ ({\cal H} {\color{blue} +\beta \dot{\phi}} )\theta_{c} = k^{2} \Psi + k^{2} \beta \delta \phi \label{eq:CQeuler_c}\\
\label{eq:dphi_cq}
&\delta \phi = - \frac{\beta \rho_c \delta_c}{m^2 + k^2/a^2} \\
\label{eq:Veff_cq}
&\square \phi = V_{, \phi}+\beta \rho_{c} \equiv V^{\rm eff},_{\phi} \\
&\ddot{\delta}+ \HH \dot{\delta} = 4 \pi G a^2 \rho\delta\left[ 1 + {2 \tilde{\beta}^2k^2\over a^2m^2 + k^2} \left( \rho_c \over \rho \right)^2 \left( \delta_c \over \delta \right) \right]
\label{eq:delta_ddot_cq}
\end{align}
%\ea
\end{minipage}
}

\vspace{0.2in}

\noindent where the over-dots denote derivatives with respect to the conformal time $\tau$, $\square=\nabla^\mu\nabla_\mu$,
${\cal H} = a^{-1}da/d\tau$, $\beta=A,_{\phi}/A$ is the scalar field coupling strength, $\tilde{\beta}^2 = \beta^2/ 8\pi G$, and $m^2=V^{\rm eff},_{\phi\phi}$,
with the effective potentials defined via Eqs.~(\ref{eq:Veff_st}) and (\ref{eq:Veff_cq}). Note that the effective potential in CQ only depends on DM. For simplicity, we assume here $A^{-2} \approx 1$ and neglect it in our equations. In the case of GBD, this implies that our $G$ is the $G$ today, while the overall change in the gravitational coupling with redshift is constrained to be very small in screened GBD theories~\cite{Wang:2012kj}. 
In the case of CQ, an $A^2 \ne 1$ would simply re-scale $\beta$ in our equations. 

We see that the Euler equation for DM in CQ, (\ref{eq:CQeuler_c}), contains a term proportional to $\dot{\phi}$, which we marked in blue. This term can be important in CQ models in which $\dot{\phi}\sim\cal{H}$~\cite{Baldi:2010pq}. It is, however, negligible in theories like the chameleon \cite{Khoury:2003aq} or the symmetron \cite{Hinterbichler:2010es} models, in which the scalar field remains near the minimum of a slowly changing effective potential. In what follows, we ignore this term for simplicity, since, for our purposes, it is sufficient to find one example where one cannot distinguish GBD from CQ. Either way, the presence of this term would not affect our arguments, as any modification of the Euler equation would yield an effective potential that is different from the true $\Psi$ if the RSD measurements are interpreted assuming an unmodified Euler equation.

One can see that in GBD theories, there is an extra term in the Poisson equation (\ref{eq:poisson_st}), and in addition the two potentials are different~(\ref{eq:slip}), $\Phi \ne \Psi$, hence $\eta \equiv \Phi/\Psi \ne 1$. One can combine Eqs.~(\ref{eq:poisson_st}), (\ref{eq:slip}) and (\ref{eq:dphi}) to write separate Poisson equations for the potential $\Psi$, which affects the motion of non-relativistic matter (through Eqs.~\eqref{eq:STeuler_b} and~\eqref{eq:STeuler_c}), and the Weyl potential $\Phi+\Psi$ felt by relativistic particles:
\ba
k^2 \Psi &=& -4 \pi Ga^2 \left[1+ {2 \tilde{\beta}^2k^2\over a^2m^2 + k^2} \right]  \left(\rho_{b} \delta_{b} + \rho_{c} \delta_{c}\right)\, , \\
k^2 (\Phi + \Psi) &=& -8 \pi G a^2 \left(\rho_{b} \delta_{b} + \rho_{c} \delta_{c}\right) \ . \label{eq:Sigma}
\ea
Comparing the above to the commonly used phenomenological parameterization of modified gravity effects on cosmological perturbations,
\ba
k^2 \Psi &=& -4 \pi \mu(a,k) G a^2 \left(\rho_{b} \delta_{b} + \rho_{c} \delta_{c}\right) \\
k^2 (\Phi + \Psi) &=& -8 \pi \Sigma(a,k) G a^2 \left(\rho_{b} \delta_{b} + \rho_{c} \delta_{c}\right) \ ,
\ea
we have 
\be
\mu=1+{2 \tilde{\beta}^2k^2\over a^2m^2 + k^2} \ , \ \ \Sigma = {1\over 2}\mu (1+\eta) = 1 \ .
\ee
Thus, GBD theories predict $\mu \ne \Sigma$. Note that this is true even if we do not assume $A^{-2} \approx 1$, in which case $\mu=A^2(1+ 2 \tilde{\beta}^2k^2 /(a^2m^2 + k^2))$ and $\Sigma=A^2$.
Moreover, we can combine the continuity and Euler equations, and use Eq.~(\ref{eq:dphi}), to derive a second order equation describing the evolution of the total matter density contrast 
$\delta = (\rho_{b} \delta_{b} + \rho_{c} \delta_{c})/(\rho_{b} + \rho_{c})$, 
given  by Eq.~(\ref{eq:delta_ddot_st}), which can be interpreted as growth in the presence of an effective gravitational coupling, $G^{\rm GBD}_{\rm eff}$, defined as
\begin{align}
\frac{G^{\rm GBD}_{\rm eff}}{G} = \mu = 1+{2 \tilde{\beta}^2k^2\over a^2m^2 + k^2} \, . \label{eq:GST}
\end{align}

In contrast, in the case of CQ, the Einstein equations are not modified and, formally, $\mu=\Sigma=\eta=1$. The effect of the scalar force on structure growth comes through the new term in the Euler equation for DM~\eqref{eq:CQeuler_c}. The second order equation for the total matter density contrast, $\delta$, in this case, is given by Eq.~(\ref{eq:delta_ddot_cq}), which can also be interpreted as growth in the presence of an effective gravitational coupling, $G^{\rm CQ}_{\rm eff}$, defined as
\be
{G^{\rm CQ}_{\rm eff} \over G} = 1 + {2 \tilde{\beta}^2k^2\over a^2m^2 + k^2} \left( \rho_c \over \rho \right)^2 \left( \delta_c \over \delta \right) \ . \label{eq:GCQ}
\ee
We see that $G^{\rm CQ}_{\rm eff}/G$ and $G^{\rm GBD}_{\rm eff}/G$ are very similar to each other. The only difference is a small suppression of the impact of the fifth force in $G^{\rm CQ}_{\rm eff}$, due to the fact that $\sim 15\%$ of matter does not feel the fifth force.

\subsection{Gravitational slip measured from galaxy peculiar velocities and weak lensing}
\label{sec:vbaryon}

The fact that $\Phi \ne \Psi$ in GBD, while $\Phi = \Psi$ in CQ, suggests that one could   differentiate the two cases by measuring $\eta$~\cite{Motta:2013cwa,Amendola:2016saw}, making it a smoking gun for modified gravity. Note that there exist scalar-tensor theories with no gravitational slip, such as Cubic Galileons~\cite{Deffayet:2009wt}, Kinetic Gravity Braiding~\cite{Deffayet:2010qz} and the ``no-slip gravity''~\cite{Linder:2018jil}, but these can be viewed as rare exceptions within the broad class of Horndeski theories~\cite{Horndeski:1974wa,Deffayet:2011gz}. 
To measure $\eta$, one can, in principle, combine weak lensing data, that measure $\Phi+\Psi$ and are, consequently, sensitive to $\Sigma$, with a measurement of the baryon velocities, that are driven by $\Psi$ and are, consequently, sensitive to $\mu$. The problem with this method is that, in CQ, the baryons too are affected by the fifth force on dark matter because they are confined in galaxies. Therefore, baryon velocities are not a true tracer of the gravitational potential $\Psi$ in this case, and using them would lead to a measured $\eta^{\rm fit} \ne 1$ even if there is no intrinsic gravitational slip.

To see this, let us start by writing the observed fluctuation in the galaxy number counts as
\begin{align}
\Delta(\bn, z)&=\delta_g-\frac{1}{\HH}\partial_r(\bV_b\cdot\bn)\, , \label{eq:Delta}
\end{align}
where $r$ is the comoving distance to the galaxies and $\bn$ is the direction of observation.
Eq.~\eqref{eq:Delta} can be Fourier transformed
\begin{align}
\Delta(\bk, z)&=b\,\delta(\bk, z)-\frac{1}{\HH}\mu_k^2\,\theta_b(\bk, z)\, , \label{eq:Deltak}
\end{align}
where $\mu_k=\hat{\bk}\cdot\bn$ is the cosine of the angle between the vector $\bk$ and the direction of observation $\bn$ (which is considered fixed in the flat-sky approximation), and $b$ is the bias. The power spectrum of $\Delta$ is then given by
\begin{align}
P^{\rm gal}(k,\mu_k,z)&=b^2P_{\delta\delta}(k,z)-\frac{2b}{\HH}\mu_k^2P_{\delta\theta_b}(k,z)+\frac{1}{\HH^2}\mu_k^4P_{\theta_b\theta_b}(k,z)\, . \label{eq:Pgalbaryon}
\end{align}
Since we are interested in the galaxy power spectrum on large scales, in the linear regime $k<k_{\rm NL}$, we need to model the correlations of the baryon velocity at those scales. For this, we split the baryon velocity into two parts: the velocity of the baryons with respect to the center of mass of the galaxy, that we call $\theta_b^{\rm loc}$, and the velocity of the center of mass of the galaxy with respect to the Hubble flow, that we call $\theta_g$:
\be
\theta_b=\theta_b^{\rm loc}+\theg\, .
\ee
In both GBD and CQ models, the velocity of the baryons with respect to the center of mass obeys
\begin{align}
\dot{\theta}^{\rm loc}_{b}+ {\cal H} \theta^{\rm loc}_{b} = k^{2} \Psi +F_{\rm int}\, ,\label{eq:Vbloc}    
\end{align}
where $F_{\rm int}$ accounts for the non-gravitational interactions affecting the motion of baryons inside the galaxy. The gravitational potential can be decomposed into a local part, due to the presence of the galaxy, and a large-scale part, due to the large-scale structure of the Universe, as shown in Fig.~1,
\begin{align}
 \Psi=\Psi^{\rm loc}+\Psi^{\rm LS}\, .   
\end{align}
Eq.~\eqref{eq:Vbloc} depends on the total gravitational potential $\Psi$. However, since the galaxy is a localised object of size that is small compared to the extent of $\Psi^{\rm LS}$, the center of mass of the galaxy and the baryons are situated at almost the same value of $\Psi^{\rm LS}$. Consequently, $\Psi^{\rm LS}$ does not impact the motion of baryons \emph{inside} the galaxy, {\it i.e.}\ \emph{with respect} to the center of mass. On the other hand, $\Psi^{\rm loc}$ varies significantly over the extent of the galaxy and does contribute to Eq.~\eqref{eq:Vbloc}. We therefore obtain
\begin{align}
\dot{\theta}^{\rm loc}_{b}+ {\cal H} \theta^{\rm loc}_{b} = k^{2} \Psi^{\rm loc} +F_{\rm int}\, .\label{eq:Vbloc2}   
\end{align}
From this equation, we see that the local velocity is uncorrelated on scales larger than the size of the galaxy. The internal forces in two different galaxies are indeed uncorrelated, and the local gravitational potentials are also uncorrelated at large distance. Therefore,
\begin{align}
P_{\theta_b^{\rm loc}\theta_b^{\rm loc}}(k,z)=0, \quad\mbox{for} \quad k\lesssim 1/s_{\rm galaxy}\, ,  
\end{align}
where $s_{\rm galaxy}$ denotes the typical size of a galaxy. As a consequence, the RSD power spectrum is only affected by the motion of the center of mass of the galaxy,
\begin{align}
P^{\rm gal}(k,\mu_k,z)&=b^2P_{\delta\delta}(k,z)-\frac{2b}{\HH}\mu_k^2P_{\delta\theta_g}(k,z)+\frac{1}{\HH^2}\mu_k^4P_{\theta_g\theta_g}(k,z)\, . \label{eq:Pgal}
\end{align}
The power spectrum can be further simplified by using that in both GBD and CQ, baryons and DM obey the continuity equation, leading to
\begin{align}
\theta_g=-\dot{\delta}=-\HH f\delta\, ,\label{eq:contg}
\end{align}
where the (total) matter growth rate is defined as
\begin{align}
f\equiv\frac{d\ln\delta}{d\ln a}\, .
\end{align}
Inserting this into Eq.~\eqref{eq:Pgal}, we obtain Eq.~\eqref{eq:Pf}. From this equation we see that the RSD power spectrum can be used to measure the growth rate $f$ and constrain $G_{\rm eff}$. Alternatively, it can also be used to probe $\Psi^{\rm LS}$. In GBD, the galaxy center of mass, $\theta_g$, obeys Eq.~\eqref{eq:VgGBD} and can therefore directly be used to reconstruct $\Psi^{\rm LS}$. In CQ however, $\theta_g$ obeys Eq.~\eqref{eq:VgCQ}, 
meaning that RSD provide a measurement of $\Psi^{\rm eff}>\Psi^{\rm LS}$ due to the fifth force. Comparing $\Psi^{\rm eff}$ with $\Phi^{\rm LS}+\Psi^{\rm LS}$ inferred from lensing would give
\begin{align}
 \frac{\Phi^{\rm LS}+\Psi^{\rm LS}}{\Psi^{\rm eff}}<\frac{\Phi^{\rm LS}+\Psi^{\rm LS}}{\Psi^{\rm LS}}=2\, ,\hspace{1cm}\mbox{leading to}\hspace{1cm}  \eta^{\rm fit}=\frac{\Phi^{\rm LS}+\Psi^{\rm LS}}{\Psi^{\rm eff}}-1<1\, ,
\end{align}
{\it i.e.}\ a detection of non-vanishing gravitational slip. Again, while we used CQ to illustrate the point, the argument holds for a general dark force.

\subsection{Galaxy distribution multipoles}
\label{sec:multipoles}

In addition to RSD, the observed fluctuation in the galaxy number counts is affected by several other distortions~\cite{Bonvin2011,Yoo:2009au,Challinor:2011bk}:
\begin{align}
\Delta^{\rm rel}(\bn, z)&=\frac{1}{\mathcal H}\partial_r\Psi+\frac{1}{\mathcal H}{\dot \bV}\cdot \mathbf n 
+\left(1-5s+\frac{5s-2}{\mathcal H r}-\frac{{\dot{\HH}}}{\mathcal H^2}+f^{\rm evol}\right)\mathbf V\cdot \mathbf n\, ,\label{eq:Delta_rel}
\end{align}
where the first term on the right hand side is the gravitational redshift that probe the true Newtonian potential $\Psi$. Note that other relativistic effects contribute to $\Delta$, like Shapiro time-delay, integrated Sachs-Wolfe, and gravitational lensing~\cite{Bonvin2011,Yoo:2009au,Challinor:2011bk}. However, these effects are 
negligible at the scales and redshifts relevant for the analyses we describe here~\cite{Jelic-Cizmek:2020pkh}. 

To separate the relativistic effects from the standard density and RSD, one can expand the power spectrum in multipoles of $\mu_k$, 
\begin{align}
P^{\rm gal}_{\rm BF}(k,\mu_k,z)=\sum_\ell P^{(\ell)}_{\rm BF}(k,z)\mathcal{L}_\ell(\mu_k)\, ,
\end{align}
where $\mathcal{L}_\ell(\mu_k)$ denotes the Legendre polynomial of order $\ell$.
Using the continuity equation~\eqref{eq:contg}, the multipoles can be written as
\begin{align}
\mbox{monopole:}\quad P_{\rm BF}^{(0)}(k,z)&=\left[b_\B b_\F +\frac{1}{3}(b_\B+b_\F)f_m+\frac{1}{5}f_m^2\right]P_{\delta\delta}(k,z)\,,\label{eq:mono}\\   
\mbox{quadrupole:}\quad P_{\rm BF}^{(2)}(k,z)&=\left[\frac{2}{3}(b_\B+b_\F) f_m +\frac 47 f_m^2\right]P_{\delta\delta}(k,z)\,, \label{eq:quad}\\
\mbox{hexadecapole:}\quad P_{\rm BF}^{(4)}(k,z)&=\frac{8}{35}f_m^2P_{\delta\delta}(k,z)\,,\label{eq:hexa}\\
\mbox{dipole:}\quad P_{\rm BF}^{(1)}(k,z)&= i\,\alpha\left(f_m,\dot{f}_m,\Theta_\B, \Theta_\F \right)\frac{\HH}{k}P_{\delta\delta}(k,z) + i(b_\B-b_\F)\frac{k}{\HH}P_{\delta\Psi}(k,z)\, ,\label{eq:dip1}\\
\mbox{octupole:}\quad P_{\rm BF}^{(3)}(k,z)&= i\,\beta\left(f_m,\Theta_\B, \Theta_\F \right)\frac{\HH}{k}P_{\delta\delta}(k,z)\, , \label{eq:oct}
\end{align}
where $\Theta_\B$ and $\Theta_\F$ encode the dependence of the multipoles on the bias, magnification bias and evolution bias of the bright and faint population, respectively. These multipoles can be measured separately by weighting the galaxy power spectrum with the appropriate Legendre polynomial
\begin{align}
P_{\rm BF}^{(\ell)}(k,z)=\frac{2\ell+1}{2}\int_{-1}^1d\mu_k\mathcal{L}_\ell(\mu_k)P^{\rm gal}_{\rm BF}(k,\mu_k,z)\, .   
\end{align}

The monopole, quadrupole and hexadecapole are routinely measured for one population of galaxies, see {\it e.g.}\ ~\cite{Alam:2020sor}, and also for multiple populations~\cite{Blake:2013nif,Zhao:2020tis}. Measuring these multipoles is actually the optimal way to extract information from RSD and to infer the growth rate $f$. Measuring the dipole is significantly more difficult, since its signal-to-noise ratio is much smaller than that of the even multipoles. This is due to the fact that the dipole is suppressed by a factor $\HH/k$ with respect to the even multipoles. Note that here we show the multipoles of the power spectrum. In practice, when including relativistic effects, it is better to work with the multipoles of the correlation function, since wide-angle corrections can be correctly accounted for in this case.

\vspace{0.3cm}

\section{Supplementary Information}

From the discussion above we see that current large-scale structure data from galaxy surveys cannot distinguish between a modified gravity of GBD type and a CQ, and, hence, distinguish a modified gravity (MG) from a dark sector force. 

As noted in the introduction, a measurement of $\eta^{\rm fit} \ne 1$ would signal new physics either way.
And, one could still learn something from measurements of $\mu^{\rm fit}$ and $\Sigma^{\rm fit}$. For instance, a measurement of $\Sigma^{\rm fit} \ne 1$ or $\Sigma^{\rm fit} = \mu^{\rm fit} \ne 1$, or $\mu^{\rm fit}<1$, would rule out both GBD and CQ. (A measurement of $\Sigma^{\rm fit} = \mu^{\rm fit}=1$ would be consistent with all mentioned theories, since they all include $\Lambda$CDM as a special case.) 
If one measures $\Sigma^{\rm fit} \ne 1$ then, in principle, any pair of values of $\Sigma^{\rm fit}$ and $\mu^{\rm fit}$ could be accommodated within a general Horndeski theory. However, as shown in~\cite{Pogosian:2016pwr,Peirone:2017ywi,Espejo:2018hxa}, one generically expects $\Sigma$ and $\mu$ to be positively correlated in Horndeski, {\it i.e.} $(\mu-1)(\Sigma-1) \ge 0$. 
The main implications of observed values of $\mu$, $\Sigma$ and $\eta$ for CQ and Horndeski are summarized in Table~\ref {tab:implications}.

\begin{table}[tbph]
\caption{Theoretical implications of observed $\mu$, $\Sigma$ and $\eta$ for CQ and modified gravity theories.}
    \label{tab:implications}
    \centering
    \begin{tabular}{c|c}
    \hline
    $\mu^{\rm fit}=\Sigma^{\rm fit}=\eta^{\rm fit}=1$  & Consistent with $\Lambda$CDM; MG and CQ constrained but not ruled out  \\
\hline    
    $\mu^{\rm fit}>1$, $\Sigma^{\rm fit}=1$ ($\eta^{\rm fit}<1$)  &  Evidence for CQ or MG, GBD unlikely due to screening constraints \\
\hline    
    $\mu^{\rm fit}<1$, $\Sigma^{\rm fit}=1$ ($\eta^{\rm fit}>1$)  &  CQ and GBD ruled out; could be other Horndeski\\
\hline    
    $\mu^{\rm fit}=\Sigma^{\rm fit} \ne 1$ ($\eta^{\rm fit}=1$)  &  CQ and GBD ruled out; could be no-slip Horndeski (CC, KGB, NSG) \\
\hline   
    $(\mu^{\rm fit}-1)(\Sigma^{\rm fit}-1) < 0$ (any $\eta^{\rm fit}$)  &  CQ and GBD ruled out; hard to accommodate with Horndeski\\
\hline 
    \end{tabular}
\end{table}

One could wonder if the problem of measuring $\Psi$ from RSD could be evaded by observing the baryon velocity field in regions where they are not bound to dark matter halos, {\it i.e.} such that $\theta_b$ would probe the large-scale gravitational potential $\Psi^{\rm LS}$. In theory, such a remote possibility could be provided by a 21\,cm map of the neutral hydrogen density during the dark ages. However, even if such observations were possible, it is likely that the hydrogen signal we would detect would come from the higher density regions that would trace the halos. Similarly, X-ray emission from ionized gas and the Sunyaev-Zeldovich contribution to CMB comes from the clustered environment in which dark matter plays a dominant role. 

Of course, if one concentrates on particular models of modified gravity or dark fifth force, there could be additional phenomenology that one could identify and constrain. While such tests are no longer model-independent, they nevertheless provide a way of ruling out large classes of possibilities. 

For example, in the case of MG theories, there can be additional observational signatures related to the type of screening mechanism they employ to stay consistent with all the stringent tests passed by GR~\cite{Will:2014kxa}. Because GR works so well inside the Solar System, any MG theory must effectively reduce to GR either on Solar System scales or at Solar System densities, a property known as screening. On the other hand, a dark force acting on DM is not subject to this requirement. For example, if one observes enhanced growth on linear scales that is consistent with both GBD and CQ, it is unlikely to be GBD, since the known ways to screen the scalar force in GBD theories is via one of the environment-dependent mechanisms, {\it e.g.}\ Dilaton~\cite{Damour:1994zq}, Cameleon~\cite{Khoury:2003aq} and Symmetron~\cite{Hinterbichler:2010es}. All of these share a similar ``thin shell'' condition, which prevents having observable effects on scales above $1$\,Mpc~\cite{Wang:2012kj,Joyce:2014kja}. Such a constraint, however, does not apply to Horndeski theories with Vainshtein-type screening \cite{Vainshtein:1972sx}. Hence, in practice, we are unlikely to confuse CQ with GBD but, instead, we could confuse it with a more general Horndeski theory.

In theories with environment-dependent screening, one can have galaxies of the same mass falling at different rates, depending on whether the matter inside them is screened or not~\cite{Hui:2009kc}. One of the manifestations of this would be a violation of a consistency relation between galaxy correlation functions, that are consequences of symmetries and hold if all galaxies experience the same gravitational potential~\cite{Kehagias:2013rpa}. Such effects would be absent in MG theories with non-environmental type screening. Hence, one cannot, in principle, differentiate between a dark force and a Horndeski theory with Vainshtein-type screening. One could indeed also have galaxies falling at different rates in dark force theories if the DM fraction of the galaxy mass varied significantly.

Other observational signatures of MG and dark forces one could look for to constrain specific models include scale-dependence of the growth factor, a negative correlation between CMB temperature and galaxies caused by the Integrated Sachs-Wolfe effect~\cite{Zucca:2019ohv} during matter dominated epoch, and new non-Gaussian signatures~\cite{Peloso:2013spa}.

In addition to large-scale structure and CMB observations, gravitational waves (GW) provide a novel avenue to test MG. For example, there is a well-known connection between the modified propagation speed of GW and the intrinsic gravitational slip~\cite{Amendola:2014wma,Saltas:2018fwy}. However, the GW speed has been strongly constrained~\cite{Lombriser:2015sxa,Amendola:2017ovw,Noller:2020afd} by simultaneous observation of the gravitational and electromagnetic radiation from merging neutron stars, GW170817~\cite{LIGOScientific:2017vwq}.
Furthermore, there are MG theories, like GBD, in which $\eta\neq 1$ but the propagation speed of GW is not modified. Hence, GW speed cannot be used to unambiguously distinguish between a dark fifth force and a modification of gravity. 

Another potentially promising way to test gravity, actively investigated in recent literature, is based on the effect of MG on the luminosity distances to GW sources~\cite{Lagos:2019kds,Dalang:2019fma,Dalang:2019rke}, where the GW data is considered in combination with electromagnetic tracers of large scale structure~\cite{Libanore:2021jqv,Scelfo:2022lsx}.

\begin{center}
{\bf Data availability statement}
\end{center}

Data sharing not applicable to this article as no datasets were generated or analysed during the current study.

\vspace{-0.5cm}

\acknowledgments 
We thank Sveva Castello, Hamid Mirpoorian, Alessandra Silvestri and Zhuangfei Wang for useful discussions, and Ruth Durrer, Kazuya Koyama and Martin Kunz for their valuable feedback on the draft of this paper. CB acknowledges financial support from the Swiss National Science Foundation and from the European Research Council (ERC) under the European Union’s Horizon 2020 research and innovation program (Grant agreement No.~863929; project title ``Testing the law of gravity with novel large-scale structure observables"). LP is supported by the National Sciences and Engineering Research Council (NSERC) of Canada.

\bibliographystyle{unsrt}	
\bibliography{dm_vs_mg.bib}
	
\end{document}